\documentclass[prd,twocolumn,showpacs,showkeys,amsmath,amssymb]{revtex4}

\usepackage{times}
\usepackage{graphicx}

\newcommand{\beq}{\begin{equation}}
\newcommand{\eeq}{\end{equation}}

\newcommand{\ba}{\begin{eqnarray}}
\newcommand{\ea}{\end{eqnarray}}

\newcommand{\bfalpha}{\mbox{\boldmath $\alpha$}}

\newcommand{\bfvt}{\mbox{\boldmath $\vartheta$}}
\newcommand{\bft}{\mbox{\boldmath $\theta$}}
\newcommand{\bfb}{\mbox{\boldmath $\beta$}}

\newcommand{\bfv}{\mathbf{v}}
\newcommand{\bfB}{\mathbf{B}}
\def\bfv{\mathbf{v}}

\def\t{\theta}

\def\ve{\varepsilon}
\def\vt{\vartheta}
\def\d{\delta}
\def\b{\beta}

\newcommand{\aap}{A\&A}
\newcommand{\mnras}{MNRAS}

\def\gs{\mathrel{\lower0.6ex\hbox{$\buildrel {\textstyle >}\over{\scriptstyle \sim}$}}}
\def\ls{\mathrel{\lower0.6ex\hbox{$\buildrel {\textstyle <}\over{\scriptstyle \sim}$}}}

\begin{document}

\title{On aberration in gravitational lensing}
\author{M. Sereno}
\email{sereno@physik.unizh.ch} 

\affiliation{Institut f\"{u}r Theoretische Physik, Universit\"{a}t Z\"{u}rich,
Winterthurerstrasse 190, CH-8057 Z\"{u}rich , Switzerland
}

\date{August 20, 2008}

\begin{abstract}
It is known that a relative translational motion between the deflector and the observer affects gravitational lensing. In this paper, a lens equation is obtained to describe such effects on actual lensing observables. Results can be easily interpreted in terms of aberration of light-rays. Both radial and transverse motions with relativistic velocities are considered. The lens equation is derived by first considering geodesic motion of photons in the rest-frame Schwarzschild spacetime of the lens, and, then, light-ray detection in the moving observer's frame. Due to the transverse motion images are displaced and distorted in the observer's celestial sphere, whereas the radial velocity along the line of sight causes an effective re-scaling of the lens mass. The Einstein ring is distorted to an ellipse whereas the caustics in the source plane are still point-like. Either for null transverse motion or up to linear order in velocities, the critical curve is still a circle with its radius corrected by a factor $(1+z_\mathrm{d})$ with respect to the static case, $z_\mathrm{d}$ being the relativistic Doppler shift of the deflector. From the observational point of view, the orbital motion of the Earth can cause potentially observable corrections of the order of the $\mu$arcsec in lensing towards the super-massive black hole at the Galactic center. On a cosmological scale, tangential peculiar velocities of cluster of galaxies bring about a typical flexion in images of background galaxies in the weak lensing regime but future measurements seem to be too much challenging.
\end{abstract}


\maketitle

\section{Introduction}

Deflectors in motion have been considered several times in gravitational lensing analyses, the main motivation being that most astrophysical lenses are actually not at rest in the observer's frame of reference. To my knowledge the first analysis of the bending angle by a moving lens was performed by \citet{py+bi93}, who applied a general technique developed to study zero geodesic in a perturbed spacetime to calculate the asymptotic behavior of a light ray deflected by a point mass moving on a Minkowski background. In the approximation of velocities much smaller than the speed of light, $v \ll c$, they obtained
\beq
\alpha \simeq -(1+z_\mathrm{d})4m/r_\mathrm{min},
\eeq
where $m$ is the mass of the lens, $r_\mathrm{min}$ the distance of closest approach of the light ray to the deflector and $z_\mathrm{d} (\sim v/c)$ is the Doppler shift of the deflector. Up to first order in $v/c$, the deflection is affected only by the component of the speed along the line of sight. The authors noted the agreement with a Lorentz transformation of the usual static bending angle to a frame in which the lens is moving. The trajectory of light rays in the field of an ensemble of moving point masses was then considered in \citet{ko+sc99}.

Later on \citet{fri+al02} confirmed that the factor $(1+z_\mathrm{d})$ corrects the bending angle with respect to the deflection angle of the same deflector at rest. The authors formulated a novel version of the Fermat principle based on envelopes of light-like geodesics and re-derived the bending angle for the case of a lens moving along the line of sight. Exploiting their method, they could also derive a lens equation in terms of distances and angles as measured in the frame relevant for observation. The deflection angle for the moving lens was still written in terms of the corresponding static angle. The $(1+z_\mathrm{d})$ correction in the case of both radial and transverse slow motion across the line of sight was later re-derived in \citet{fri03a}, who integrated the zero geodesics in presence of a time-dependent perturbation keeping terms up to linear order in velocity. Such analyses on the bending angle were eventually extended to arbitrary large radial velocities \citep{wu+sp04}, uniformly moving monopoles \citep{kli03}, binary systems \citep{hey05} and extended mass distributions \citep{ser05}. 

From the above review, a clear understanding of how a relative motion between the lens and the observer affects the bending angle emerges. The picture is in full agreement with standard aberration, according to which angles made by light rays as observed in a moving frame are corrected by a factor $(1+z_\mathrm{d})$ with respect to angles made by the same photons as observed in the rest frame. Such a correction can be also interpreted in terms of standard aberration of light propagating within an optically active medium with effective index of refraction induced by the gravitational field of the lens in motion \citep{fri03b}. Furthermore, the different nature between the effect of a translational motion, related to Lorentz invariance, and a non null angular momentum, connected to frame-dragging has been also well understood \citep[and references therein]{ser05}. 

What is still missing is a complete description of the effect of a relative translational motion on actual lensing observations. The bending angle is defined as the difference between the asymptotic directions of the light ray at the source and at the observer and is usually calculated as a coordinate difference in a suitably defined system. On the other hand, observers measure image positions in the celestial sphere rather than bending angles at the lens position. The aim of the present paper is to derive a lens equation for the observed image angular positions in a consistent way and without referring to a corresponding static angle. In general, the bending angle for lens and observer in a static configuration can be written in terms of the rest mass of the lens and of the impact parameter of the light ray, which in turn is related to the constants of motion of the light-like geodesics. The approach proposed here avoids problems connected to the correction of such a static angle to the moving frame by considering the geodesic motion of the light rays in the rest frame of the lens. Assuming the lens to be point-like, this is the well known problem of null geodesics in the Schwarzschild metric. The observer will be moving in this space-time. To account for the relative motion between lens and observer, the components of the four-momentum of the photons reaching the detector will be then evaluated in the moving frame and translated in observed angular positions in the observer's sky. This approach allows an easy comparison of the outputs of measurement processes for lens and observer which either stay put or have a relative motion. Furthermore, it allows to extend previous analyses to relativistic velocities for transverse as well radial motions. Calculations will be limited to the first order in deflection for photon in the weak deflection limit.

In addition to clarify some theoretical aspects, the second main motivation to study lensing observables by moving deflectors is the potential astrophysical relevance. The supermassive black hole hosted in the radio source Sgr~A* in the Galactic center has emerged as an appealing target for testing higher order effects in gravitational lensing with future space- and ground-based experiments \cite[and references therein]{se+de07} and demands a full theoretical understanding of all effects at least at the level of the $\mu$arcsecond. Furthermore deflectors in the Milky Way may have quite large velocities, whereas, on a cosmological scale, galaxies and galaxy clusters move with quite large peculiar velocities out of the Hubble flow. Finally, there has been some evidence that correction factors in the lensing  time delay of order of $v/c$ have been actually measured at the time of a close alignment of Jupiter with the quasar J0842+1835 \citep{fo+ko03}.

The paper is organized as follows. In Section~\ref{sec:geod}, we review the light-like geodesic motion in a Schwarzschild metric and the properties of the four-momentum of the photons in the locally flat frame of a moving observer. In Section~\ref{sec:lens}, the lens equation is derived and solved in the first order of deflection. The effect of aberration on the Einstein ring is discussed in Section~\ref{sec:eins} together with image distortion. Section~\ref{sec:radi} is dedicated to the case of pure radial motion whereas Section~\ref{sec:astr} reviews the order of magnitude of the effect in astrophysical lensing system on several scales. Section~\ref{sec:conc} is devoted to some final considerations and comments.

\section{Geodesic motion and observer's frame}
\label{sec:geod}

The lens equations are a mapping between the angular position of the source and the angular position of the images in the observer's sky and can be derived from the geodesic equations. Since we are considering a static spherically symmetric lens and a moving observer, the space-time in the vacuum region outside the lens can be described by the Schwarzschild metric. In Boyer-Lindquist coordinates,
\beq
\label{sds1}
ds^2= f(r;m) dt^2- f(r;m)^{-1 } dr^2 -r^2 \left( d \theta^2- \sin^2 \theta d\phi^2 \right),
\eeq
where $m$ is the black hole mass and
\beq
f(r;m) \equiv 1- \frac{2 m}{r} .
\eeq
 We are using units $G=c=1$. The polar variable can be rewritten as $\mu \equiv \cos \theta$. 

Photon trajectories go from a source $S$ in $\{r_\mathrm{s}, \theta_\mathrm{s}, \phi_\mathrm{s}\}$ to an observer $O$ in $\{r_\mathrm{o}, \theta_\mathrm{o}, \phi_\mathrm{o}=0\}$. Both $S$ and $O$ are located in the nearly flat region of the spacetime, very far from the lens.  Lensing in a Schwarzschild space-time is spherically symmetric and analyses are usually restricted to the equatorial plane. However, since the proper motion of the observer breaks the symmetry, the geodesic equations must be considered in their general form. 
Light ray motion in the exterior field of a black hole is a well known problem and has been throughly studied even in presence of a non null angular momentum \citep{car68,cha83}. One simple way to build on previous results is to consider the results for the geodesic motion in the Kerr metric and then putting the spin parameter to zero. Without loss of generality, we can also fix the polar coordinate of the observer, $\mu_\mathrm{o}=0$. 

The geodesics for a light ray can be expressed in terms of the first integrals of motion $J$ and $Q$, which are strictly related to the image positions \citep{car68,cha83}. Our interest is on lensing in the weak deflection limit, so that light-like geodesics can be suitably expanded in the small parameter $\epsilon \sim m/b $ with $b \equiv \sqrt{J^2 +Q} $ \citep{bra86,se+de06,se+de07}.  In asymptotic nearly flat regions, $b /r_\mathrm{o} \sim b /r_\mathrm{s} \sim \epsilon$. Once the expansion in $\epsilon$ is performed up to and including terms of order of ${\cal{O}}(\epsilon^2)$, the geodesic relations take the form
\begin{eqnarray}
\phi_s  &  \simeq  & -\pi \frac{b_1}{|b_1|} -\frac{4m}{b}\frac{b_1}{b} + b_1 \left(\frac{1}{r_\mathrm{s}}+\frac{1}{r_\mathrm{o}}\right)- \frac{15 \pi}{4}  \frac{m^2}{b^2} \frac{b_1}{b} , \label{geod1}  \\
\mu_s  &  \simeq  & -\frac{4m}{b}\frac{b_2}{b} + b_2 \left(\frac{1}{r_\mathrm{s} }+\frac{1}{r_\mathrm{o}}\right)  - \frac{15 \pi}{4}  \frac{m^2}{b^2} \frac{b_2}{b}. \label{geod2}
\end{eqnarray}
The parameters  $b_1$ and $b_2$ accounts for the impact vector and are convenient rewritings of the usual constants of motion $J$ and $Q$ \citep{se+de06}. For an equatorial observer, $\mu_\mathrm{o}=0$,
\ba
b_1   &  \equiv &  - J, \\
b_2   &  \equiv &  - (-1)^k \sqrt{Q} .  
\ea
The parameter $k$ is an even (odd) integer for photons coming from below (above). Details can be found in \citet{se+de06}.

The angular position of the images in the observer's sky depends on the constants of motion. The frame of reference of the moving observer can be oriented parallely to the local flat three-space of the static frame centered at the same coordinates. Position angles of the images in the observer's sky can be expressed in terms of the tetrad components of the four momentum $P$ of the photon \citep{bar+al72,vie93}. Let the moving observer have a four velocity $U = U^t (1, v^r, v^\t, v^\phi)$ in Boyer-Lindquist coordinates. In the tetrad frame of the static observer, the components of the three velocity of the observer are $\bfv \equiv \{ v^{(r)}, v^{(\t)}, v^{(\phi)}\}$. Note that the velocity components in the tetrad system, $v^{(i)}$, have units of meters per second once the physical units are restored. Once we have the components of the four-momentum $P$ of the light in the static tetrad systems, the components in the moving frame can be calculated through a local Lorentz transformation with relative velocity $\bfv$. The angles $\vt_1$ and $\vt_2$ in the observer's sky, denoting the angular distance from the coordinate axes, are then defined such that $\tan \vt_1 = -P^{[\phi]}/ P^{[r]}$ and $\tan \vt_2 = P^{[\theta]}/ P^{[r]}$, with the square brackets denoting components in the moving tetrad frame. The components of the four momentum can be written as
\begin{widetext}
\ba
P^{[t]} & = & \frac{\gamma}{\sqrt{f(r_\mathrm{o};m) }}
 \left[
       1- v^{(r)} \sqrt{1- \left( \frac{b}{r_\mathrm{o}}\right)^2  f(r_\mathrm{o};m)} 
      +    \sqrt{f(r_\mathrm{o};m) } 
        \left( v^{(\theta)} \frac{b_2}{r_\mathrm{o}}  + v^{(\phi)} \frac{b_1}{r_\mathrm{o}}  \right)
\right]  , \\
P^{[\phi]}        & = &  -\frac{b_1}{r_\mathrm{o}} - \frac{\gamma  v^{(\phi)} }{ \sqrt{f(r_\mathrm{o};m)}  }   
\left[
1  -   \frac{ \gamma v^{(r)} }{1 + \gamma}   \sqrt{ 1- \left( \frac{b}{r_\mathrm{o}} \right)^2   f(r_\mathrm{o};m) } 
 +  \frac{\gamma}{1 + \gamma}   \sqrt{f(r_\mathrm{o};m)}     \left( v^{(\phi)} \frac{b_1}{r_\mathrm{o} }  + v^{(\theta)} \frac{b_2}{r_\mathrm{o}} \right)
 \right]   , \\
P^{[r]}  & = &    \frac{\gamma}{\sqrt{f(r_\mathrm{o};m)}} \left\{\left(\frac{\gamma   {v^{(r)}}^2}{\gamma +1}+\frac{1}{\gamma }\right)
   \sqrt{1-\left( \frac{b}{r_\mathrm{o}}\right)^2 f(r_\mathrm{o};m)}  
    - v^{(r)} \left[   1+ \frac{\gamma}{1+\gamma }  \sqrt{f(r_\mathrm{o};m)} \left(
           v^{(\theta)}  \frac{b_2}{r_\mathrm{o}}  +   v^{(\phi)} \frac{b_1}{r_\mathrm{o}} \right)
       \right] \right\}   , \\
P^{[\t]} & = & -\frac{b_2}{r_\mathrm{o}} - \frac{\gamma  v^{(\t)} }{ \sqrt{f(r_\mathrm{o};m)}  }   
\left[
1  -   \frac{ \gamma v^{(r)} }{1 + \gamma}   \sqrt{ 1- \left( \frac{b}{r_\mathrm{o}} \right)^2   f(r_\mathrm{o};m) }   
+  \frac{\gamma}{1 + \gamma}  \sqrt{f(r_\mathrm{o};m)}      \left( v^{(\phi)} \frac{b_1}{r_\mathrm{o} }  + v^{(\theta)} \frac{b_2}{r_\mathrm{o}} \right)
 \right]  ,
\ea
\end{widetext}
where $\gamma \equiv  1/\sqrt{1-v^2}$, with $v = \sqrt{  {v^{(r)}}^2+{v^{(\phi)}}^2+ {v^{(\theta)}}^2  }$. The size of the transverse motion across the line of size will be expressed by means of the perpendicular velocity $v_\perp = \sqrt{ {v^{(\phi)}}^2+ {v^{(\theta)}}^2  }$.
 
Due to aberration, the angular position of the lens is shifted from the coordinate centre in the observer's sky. The un-deflected light ray from the lens to the observer, which identifies the line of sight, is picked up by choosing the parameters $b_1=b_2=0$. Then, the angular coordinates of the lens in the tetrad system of the moving observer read
\ba
\tan \vt_{1}^\mathrm{los} & = &
\frac{v^{(\phi)} }{\frac{1}{\gamma }-v^{(r)}+\frac{\gamma   {v^{(r)}}^2}{\gamma +1}} \left(1-\frac{v^{(r)} \gamma }{\gamma +1}\right) ,\\
\tan \vt_{2}^\mathrm{los} & = &
-\frac{v^{\theta }}{\frac{1}{\gamma }-v^{(r)}+\frac{\gamma   {v^{(r)}}^2}{\gamma +1}}  \left(1-\frac{v^{(r)} \gamma }{\gamma +1}\right).
\ea
The above expressions are full consistent with the standard aberration formula \citep[and references therein]{fri03b}, in which, at linear order in velocities, the transverse component of the velocity adds a fixed shift to each small angle. In the next section, we will be mainly concerned with angular displacement from the line of sight, rather than `absolute' angular positions. 

The `unlensed' source position can be written in terms of the angular position $\bfB = \{ B_1, B_2\}$ at which the source would be seen by the observer in absence of the lens, i.e for $m=0$; $\bfB$ is then given in terms of fictitious constants of motion which solve the geodesic motion for the actual source and observer coordinates but for $m=0$ \citep{ser08}.  The unlensed four-momentum associated to the angular position $\bfB$ will be then given in terms of such fictitious constants.

\section{Lens equation}
\label{sec:lens}

To get the lens equation in terms of the angular position of the unlensed source, $\bfB$, and of the observed images, $\bfvt$, we can proceed as follows \citep{ser08}.  We write the source azimuthal, $\phi_\mathrm{s}$, and polar coordinate, $\mu_\mathrm{s}$, first in terms of $\bfB$, plugging the ``unlensed'' constants of motion in the geodesic equations, then in term of the image positions, plugging in the actual impact parameters written in term of the observed angle $\bfvt$. The lens equations are finally obtained by equating the corresponding expressions,
\ba
\phi_\mathrm{s}(\bfvt; m) & = & \phi_\mathrm{s} (\bfB; m=0), \label{lens1} \\
\mu_\mathrm{s}(\bfvt; m) & = & \mu_\mathrm{s} (\bfB; m=0). \label{lens2} 
\ea
Such a procedure is general and would provide exact relations if applied to the geodesic equations in their integral form. However, we are interested on the main effect of aberration on lensing observations, so that we can consider the geodesic equations in their expanded form, Eqs.~(\ref{geod1},~\ref{geod2}). To consider aberration, it is enough to consider the first order in the deflection, i.e. to take terms in the expansion up to and including terms of order of ${\cal{O}}(\epsilon)$, which account for the first order perturbation with respect to the undeflected path.

As usual when using angular coordinates instead on the invariants of motion, it can be appropriate to introduce a series expansion parameter in the weak deflection limit based on the angular Einstein ring. In terms of radial coordinates, the Einstein radius for a static observer is defined as \citep{se+de06}, 
\beq
\label{lens3}
\vt_\mathrm{E} \equiv \sqrt{4 m \frac{r_\mathrm{s}}{r_\mathrm{o}(r_\mathrm{o}+r_\mathrm{s})}} ;
\eeq
the expansion parameter $\ve$ is then defined as $\varepsilon  \equiv  \vt_\mathrm{E}/4 D$, where $D \equiv r_\mathrm{s}/(r_\mathrm{o}+r_\mathrm{s})$ \citep{ke+pe05,se+de06}.  In the background Minkowski space ($m=0$), radial coordinates can be identified with angular diameter distances as measured from a static observer. For lens and source aligned with the line of sight, the distance from the observer to the lens is $D_\mathrm{d}^0 = r_\mathrm{o}$ and the distance from the lens to the source is $D_\mathrm{ds}^0 = r_\mathrm{s}$. In the Euclidean space ($m=0$), the distance from the static observer to the source can be written as $D_\mathrm{s}^0 =D_\mathrm{d}^0 + D_\mathrm{ds}^0 = r_\mathrm{o}+r_\mathrm{s}$. Note that we are considering a Minkowski background whereas in a more general cosmological framework one should consider the Roberton-Walker spacetime. However, the expressions for the Einstein ring and other lensing quantities can be generalized by a proper use of angular diameter distances, written in the expanding space, instead of radial positions.

We assume that up to first order in $\ve$  the image positions can be written as,
\beq
\label{lens4}
\bfvt  \simeq  \bfvt^\mathrm{los} + \vt_\mathrm{E} \delta \bft,
\eeq
where $\delta \bft $ accounts for the angular separation in the lens plane between the image position and the line of sight. The displacement of the source from the line of sight can be rescaled as well in terms of the Einstein radius as $\delta \bfb = (\bfB - \bfvt^\mathrm{los}) / \vt_\mathrm{E}$.  At first order in deflection, the lens equation takes the form
\beq
\label{lens5}
\delta \bfb   = \delta \bft \left(  1 -\frac{A_\mathrm{E}^2}{\vt^2_\mathrm{ell}} \right), 
\eeq
where $\vt^2_\mathrm{ell}$ is an elliptical radius in the lens plane,
\beq
\label{lens6}
\vt^2_\mathrm{ell} \equiv A_{11} \delta \t_1^2 + 2 A_{12} \delta \t_1 \delta \t_2 +A_{22} \delta \t_2^2,
\eeq
with
\ba
A_{11}   &  \equiv & 1 -A_a \left( {v^{(\t)}} \right),  \label{lens7} \\
A_{12}   &  \equiv & A_a \left( \sqrt{  v^{(\theta)} v^{(\phi)}  }  \right) , \label{lens8}  \\
A_{22}   &  \equiv & 1 -A_a \left(  {v^{(\phi)}}  \right), \label{lens9}
\ea
and
\beq
\label{lens10}
A_a (x) \equiv  \frac{x^2}{\left( 1-v^{(r)}\right)^2}   \left(1- v^{(r)} \frac{ \gamma }{1+ \gamma }\right)^2.
\eeq
The factor $A_\mathrm{E}$ defines an overall rescaling of the static Einstein radius, 
\ba
\label{lens11}
A_\mathrm{E}  & =  &\frac{1+  \gamma  \left(1-v^{(r)}\right) \left(1-v^{(r)} \gamma \right)}{\left(1-v^{(r)}\right)^2 \gamma ^2 (\gamma
   +1)} \\
   & \times & \frac{1}{\sqrt{(1- A_a(v^{(\t)}))(1-A_a(v^{(\phi)})) }} . \nonumber
\ea
It is usually stated that, up to linear order in the velocities, the relative motion of the lens across the line of sight has no effect on the bending angle as long as the motion is approximately uniform \citep{fri03a,py+bi93}. This is mainly motivated by the fact that even if there is a tangential motion, the bending angle by a moving deflector can be written in terms of the same deflector at rest as $\bfalpha = (1+z_\mathrm{d}) \bfalpha_s$. However, the definition of the corresponding static angle $\bfalpha_s$ can be somewhat tricky, as it has been also shown that the transverse motion affects the impact parameter \citep{fri03a}. The method developed above avoids such problems in that it does not refer to whatever `corresponding' static bending angle. Due to a transverse motion, the line of sight is displaced from the coordinate centre. Counting angular separations from such a displaced position accounts for this effect but we can see from Eqs.~(\ref{lens5}-\ref{lens11}) that the components of the transverse velocity also enter through the elliptical radius.

\begin{figure}
        \resizebox{\hsize}{!}{\includegraphics{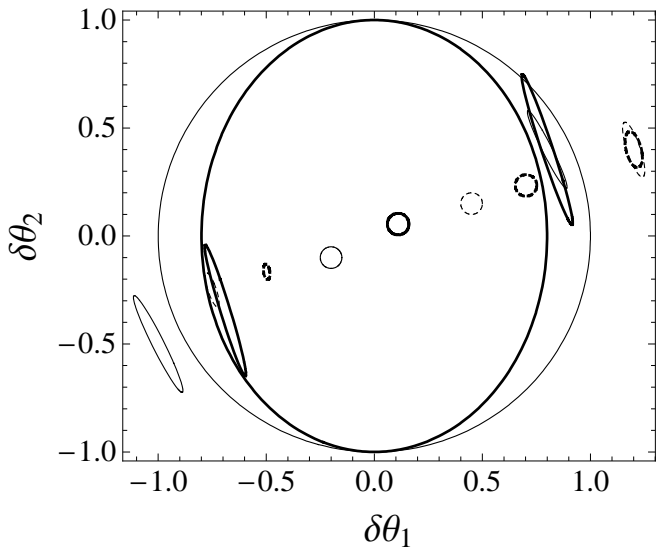}}
        \caption{Images of circular sources behind a lens moving across the line of sight. The coordinate deviations $\d \vt_1$ and $\delta \vt_2$ from the line of sight are in units of the Einstein radius for a static observer. Thick and thin lines correspond to either a moving ($v^{(r)}=0$, $v^{(\t)}=0.6$ and $v^{(\phi)} = 0$) or a static observer. Two sources behind the moving lens (full and dashed line) are considered. For each source, the corresponding source in the static case that would produce an arc at the same angular position is found and plotted. The large thick and thin curves are the critical curves for the moving and the static case, respectively. Coordinate displacements are measured from the angular position of the lens in both cases.} 
        \label{fig_imm_dist}
\end{figure}

Two images of a single source are formed. Their angular positions can be written as
\beq
\label{lens12}
\d \bft^{\pm}  =  \frac{1}{2} \left(1 \pm \sqrt{1 + \frac{4 A_\mathrm{E}^2}{\d \beta_\mathrm{ell} ^2}} \right) \d \bfb,
\eeq
with $\b_\mathrm{ell}$ defining an elliptical radius in the source plane,
\beq
\label{lens13}
\b^2_\mathrm{ell} \equiv A_{11} \delta \b_1^2 + 2 A_{12} \delta \b_1 \delta \b_2 +A_{22} \delta \b_2^2.
\eeq
Up to this order in deflection, the two images are still aligned with the lens. For small velocities
\ba
\d \bft^{\pm}  & \simeq & \frac{\d \bfb}{\d \b} \left\{ \frac{ \delta \beta}{2} \left(1 \pm \sqrt{1+\frac{4}{\delta \beta ^2}} \right) \right.   \label{lens14} \\
&  \pm &  \frac{2 v^{(r)}}{\delta \beta \sqrt{1+\frac{4}{\delta  \beta ^2} }  }   \pm   \frac{2}{ \delta \beta \sqrt{1+\frac{4}{\delta \beta ^2}} }  \nonumber  \\
& \times &
\left.  \left[    {v^{(r)}}^2  \frac{2+ \delta \beta ^2}{4 + \delta \beta^2} -\frac{1}{2} \left( v^{(\theta)}  \frac{\delta \beta _2 }{\delta \beta }+v^{(\phi)} \frac{\delta \beta _1 }{\delta \beta} \right)^2
  \right] \right\} . \nonumber
\ea
Up to linear order in the velocities, only the radial velocity enters and an angular re-scaling occurs. Circular symmetry still holds. Spherical symmetry is broken due to transverse motion, which shows up at the next order.  Given two observers, a static and a moving one, a source with the same angular position with respect to the line of sight for the two observers will form images in different positions in the celestial sphere. Stated the other way, an image in the same position as seen from either a static or a moving observer corresponds to different source positions, see Fig.~\ref{fig_imm_dist}.  From an observational point of view, the two cases can be distinguished in principle by considering the position of the counter-image. The angular splitting between the two images reads
\ba
\Delta \t &  = & \d \b  \sqrt{1 + \frac{4 A_\mathrm{E}^2}{\d \beta_\mathrm{ell} ^2}}  \label{lens15} \\
             & \simeq & \sqrt{\delta \beta ^2+4} +\frac{4 v^{(r)}}{\sqrt{\delta \beta ^2+4}}    +\frac{2}{\sqrt{\delta \beta ^2+4}}             \label{lens16} \\
             & \times  &  \left[   2 {v^{(r)}}^2  \frac{2+ \delta \beta ^2}{4 + \delta \beta^2} - \left( v^{(\theta)}  \frac{\delta \beta _2 }{\delta \beta }+v^{(\phi)} \frac{\delta \beta _1 }{\delta \beta} \right)^2 .
  \right]    \nonumber
\ea
The transverse motion affects the angular splitting only in terms quadratic in the velocities.

\begin{figure}
        \resizebox{\hsize}{!}{\includegraphics{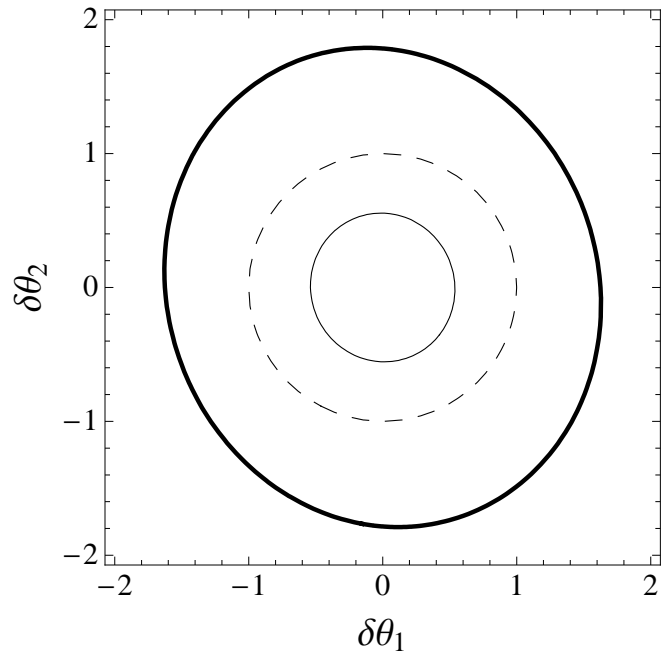}}
        \caption{Critical curves in the sky of an observer moving with relativistic velocity. The coordinate deviations $\d \vt_1$ and $\delta \vt_2$ from the line of sight are in units of the Einstein radius for a static observer. The thick and thin full curves correspond to $v^{(r)} =0.5$ and $-0.5$, respectively; the remaining components of the three velocity are fixed to $v^{(\t)}=0.1$ and $v^{(\phi)} = 0.3$. The dashed line is the Einstein circle that would be seen by a static observer. The radial velocity causes the curve to either shrink or inflate; the transverse motion determines the ellipticity and the orientation.} 
        \label{fig_curv_crit}
\end{figure}

\section{Amplification and aberrated Einstein ring}
\label{sec:eins}

It is known from the very beginning of gravitational lensing studies that a source aligned with the line of sight behind a spherically symmetric lens would be imaged in the sky of a static observer as a circle, the so-called Einstein or Chwolson ring. The luminous amplification of a point-like source in such a configuration would formally diverge. Aberration effects in the frame of a moving observer cause distortion effects. In general, the (signed) geometric luminous amplification of the image, $A$, is given by the ratio between the angular area of the image in the observer sky and the angular area of the source in absence of lensing. Up to first order in deflection,  $A$ can be calculated as the inverse of the Jacobian determinant of the lensing mapping, $J$, 
\ba
A & = & J^{-1} \label{eins1}\\
 & = & \left[ \frac{\partial B_1 \partial B_2}{\partial \vt_1 \partial \vt_2} \right]^{-1}  \label{eins2} .
\ea
Given the lens mapping in Eq.~(\ref{lens5}), the Jacobian determinant can be expressed as
\beq
\label{eins3}
J \simeq 1-\left( \frac{A_\mathrm{E}}{ \d \t_\mathrm{ell}}\right)^2.
\eeq
It is immediate to see that the locus of critical points in the lens plane is given by the relation,
\beq
\label{eins4}
 \d \t_\mathrm{ell}^2 = A_\mathrm{E}^2,
\eeq
which defines an ellipse in the observer's sky. The corresponding locations in the source plane give the caustics. Even in the case of an aberrated point-like lens, the caustics are still point-like and aligned with the line of sight to the lens, $\d \bfb =0$. The case of elliptical lenses is quite different from aberrated lensing.  Differently from deflector with either elliptical mass distribution or deflection potential  \citep{sef}, in our case the ellipticity enters the lens equations only through the elliptical radius. The transverse motion of the observer breaks the circular symmetry for images in the lens plane, but the spherical symmetry of the lens still shows up in the source plane. That is why the caustics for a moving spherical lens are still point-like whereas the caustics behind an elliptical lens get a finite astroid shape. 

The axial ratio, the semi-major axis and the angle between the major axis and the abscissa axis of the distorted Einstein ring in the observer's sky are given by, respectively,
\ba
e & = &
\frac{\left(1-v^{(r)}\right) \gamma  \left(1-v^{(r)} \gamma \right)+1}{\left(1-v^{(r)}\right) \gamma  (\gamma +1)} \\
   & \simeq & 1  -\frac{v_\perp^2}{2}    + \frac{v_\perp^2}{2}v^{(r)} ,  \\
A_{+} & = & A_\mathrm{E} \\
  & \simeq & 1 +v^{(r)}+\frac{ {v^{(r)}}^2}{2}  + \frac{v^2}{2} v^{(r)} , \\
\tan \varphi_+ & = &\frac{v^{(\t)}}{v^{(\phi)}} .
\ea
The ellipticity is caused by the component of the velocity perpendicular to the line of sight whereas the relative amplitudes of $v^{(\t)}$ and $v^{(\phi)}$ determine the orientation in the sky. The deviations from circular symmetry are quadratic in the transverse motion. In Fig.~\ref{fig_curv_crit}, the different effects of the three-velocity components are shown. The area enclosed by the critical curve grows up for a receding lens ($v^{(r)}>0$ i.e. $z_\mathrm{d}>0$) and shrinks for an approaching deflector. Up to linear order in the velocities, the critical curve is still a ring with radius rescaled by a factor $ (1+v^{(r)})$.
For a pure radial motion, the critical curve is a circle with radius
\ba
\vt_t  & =  &\vt_\mathrm{E} \sqrt{\frac{1+v^{(r)}}{1-v^{(r)}}} \\
	& \simeq & \vt_\mathrm{E} \left(1+v^{(r)} +\frac{{v^{(r)}}^2}{2} +\frac{{v^{(r)}}^3}{2} \right) .
\ea
For a radial motion, the scaling factor for the Einstein ring is then exactly $1+ z_\mathrm{d}$, being $z_\mathrm{d}$ the special relativistic Doppler redshift of the lens as measured by the observer, $1+z_\mathrm{d} = \sqrt{(1+v^{(r)})/(1-v^{(r)})}$. Such a re-scaling of the radius of the critical curve corrects previous results discussed in literature \citep{fri03a,hey05,ser05} and is one of the main result of this paper. This re-scaling is in full agreement with the aberration effect suffered by light rays, according to which angles observed in a moving frame are corrected by a factor $(1+z_\mathrm{d})$ with respect to the angles made by the same photons as observed in the rest frame.

Due to the transverse motion, the Jacobian matrix is no more symmetric so that the deflection angle in the lens equation can no more be expressed in terms of the gradient of a deflection potential. The shape of ordinary images of circular sources is no more elliptical, with relative deviations of the order of $v_\perp^2$. Images of circular sources distorted by the motion across the line of sight are plotted in Fig.~\ref{fig_imm_dist}, where lensing as seen by a static or a moving observer is compared by considering  images centered in the same angular position with respect to the lens but corresponding to different source positions for the two observers. Counter-images are no more coincident.  In circular symmetric lenses, shear causes images to be elongated tangentially with respect to the lens center. In the weak lensing regime, such tangential alignment defines on average a circular pattern. Due to the transverse motion, circle in the observer's sky changes to ellipse and images align themselves tangentially with respect to ellipses rather than circles. Aberrated images are then flexed with respect to the usual elliptical shape.

\section{Radial motion}
\label{sec:radi}

Comparison with previous works and understanding of what in case has been overlooked can be most easily performed considering pure radial motion of the source. In this case, spherically symmetry is preserved and lens equation reduces to
\ba
B  &  =         & \vt -\frac{D_\mathrm{ds}^0}{D_\mathrm{s}^0} \frac{1+v^{(r)}}{1-v^{(r)}}\frac{4m}{D_\mathrm{d}^0 \vt}, \label{radi1} \\
     &  \simeq & \vt -\frac{D_\mathrm{ds}^0}{D_\mathrm{s}^0} (1+2 v^{(r)}) \frac{4m}{D_\mathrm{d}^0 \vt}, \label{radi2}
\ea
with distances referring to a static observer. Let us introduce angular diameter distances as measured from the moving observer. The velocity of the observer affects the measurement of solid angles so that
\ba
D_\mathrm{d}     & = & \frac{D_\mathrm{d}^0}{1+z_\mathrm{d}} \\
D_\mathrm{ds} & = & D_\mathrm{ds}^0 \\
D_\mathrm{s}     & = & \frac{D_\mathrm{s}^0}{1+z_\mathrm{d}},
\ea
with $z_\mathrm{d}$ being the relativistic Doppler redshift of the lens as measures by the observer. Here, the source velocity does not play a role. Using distances corrected for the relative motion, the lens equation can be written as
\beq
\label{radi5}
B = \vt -\frac{D_\mathrm{ds}}{D_\mathrm{s}} \frac{4m}{D_\mathrm{d} \vt}.
\eeq
The lens equation written in terms of distances and angles relative to the frame of the moving observer takes the same formal expression as the classic lens equation for a static observer, as already pointed out in \citet{fri+al02}, which considered the opposite point of view of a moving lens and a static observer. It is remarkable that, in gravitational lensing phenomena, the angular diameter distances can embody most of what is connected with the background, from the effect of a relative motion in a Minkowski spacetime \citep{fri+al02}, as we have just seen, to the presence of a cosmological constant in a de-Sitter background \citep{ser08}.

However, thinking about lensing by a deflector which is moving with respect to the observer in terms of a corrected static angle $\alpha_0$ can be misleading. It has been shown with different approaches that, due to aberration, the bending angle by a moving deflector carries a pre-factor of $\sim (1+z_\mathrm{d})$ with respect to the bending angle by the same deflector at rest \citep{py+bi93,fri03a,fri03b,fri+al02}. This result has been then erroneously translated in a lens equation which is the same as that for a static configuration apart for the same pre-factor of $\sim (1+z_\mathrm{d})$ in front of the deflection angle. Such a lens equation would erroneously imply an Einstein ring corrected by a factor $\sim (1+ z_\mathrm{d}/2)$ with respect to that which shows up in the sky of a static observer \citep{fri03b,hey05,ser05}. However, the same considerations made about the effect of aberration on the deflection angle must be applied as well to the Einstein ring and to the image angular positions, which are the effective observable quantities in lensing. Then, the Einstein radius has to carry a pre-factor of $\sim (1+z_\mathrm{d})$ with respect to the circle observed in the case of a static configuration. This is implemented in Eqs.~(\ref{radi1}) which was derived first calculating the deflection angle, i.e. the coordinate variation due to lensing, in the static frame of the lens and then translating the constants of motion to angles observed in the moving frame. For small velocities, the pre-factor in Eq.~(\ref{radi2}) is then $\sim (1+2 z_\mathrm{d})$ which allows to re-scale observed angles by $\sim (1+z_\mathrm{d})$. This is consistent with the fact that in lensing the deflection angle scales as the mass $m$, whereas the observed image separation scales as the Einstein radius, i.e. as the square root of the mass.

\section{Astrophysical systems}
\label{sec:astr}

Let us consider how aberration affects some astrophysical lenses. Observations of lensing phenomena towards the super-massive black hole Sgr~A* in the Galactic center, with a mass of $\sim 3.6\times 10^6 M_\odot$ and at a distance of $\sim 7.6~\mathrm{kpc}$ from the Earth \citep{eis+al05},  performed with accuracies at the level of $\sim 1~\mu$arcsec, which are within the reach of future missions, should be able to provide important tests of general relativity.  For a source $\sim 1~\mathrm{pc}$ behind the black hole, the unperturbed Einstein ring has a radius of $\sim 0.02$~arcsec.  Then, a correction of the order of the $10^{-2}$ percent, corresponding to a radial velocity of $\sim 30~\mathrm{km}~\mathrm{s}^{-1}$ would still produce a potentially observable angular shift of the order of the $\mu$arcsec. The Earth orbits around the Sun with a mean orbital velocity of $\sim 30~\mathrm{km}~\mathrm{s}^{-1}$, so that aberration effects will play a role in accurate modelling of future lensing events by Sgr A*. 

Due to a transverse motion $v_\perp$, the circular symmetry of the Einstein ring is broken. As we have seen in the previous section, effects are quadratic in the velocity perpendicular to the line of sight and deviations from circular symmetry in the angular pattern of the images are of order of $\sim \vt_\mathrm{E} (v_\perp/c)^2$. Together with the Sun, the Earth takes part in the overall rotation of the Milky Way, which at the Sun position has a circular velocity of $\sim 300~\mathrm{km}~\mathrm{s}^{-1}$. Then, the effect on lensing by Sgr A* is $\sim 0.02$~$\mu$arcsec. On the other hand, the residual proper motion of Sgr A* perpendicular to the plane of the Galaxy is $-0.4\pm 0.9~\mathrm{km}~\mathrm{s}^{-1}$ \citep{re+br04}, too much small to give any sizable effect. 

The correction to the microlensing light curves of stars in the Galactic bulge or the Magellanic clouds has been also considered, pointing out that a study of the translational effect should require an unambiguous determination of the lensing mass together with estimates of proper motion and distance of the lensing star \citep{fri03a}.

The effect of translational motion on lensing by galaxy clusters can be more sizeable. Even if the calculations of the previous sections have been performed considering lenses in an otherwise flat Minkowski background, which is appropriate for lensing in local systems, we can still get an estimate of the effects of aberration for cosmological lenses. The idea is to use the same correction factors, which are written in terms of velocity components, being careful to re-scale angular positions in terms of an Einstein ring properly written in terms of angular diameter distances. As we have seen, a radial motion can be interpreted in terms of a re-scaling of the mass. A typical peculiar motion with respect to the Hubble flow brings about a systematic error $\ls 0.3$ per cent, independent of the mass of the cluster \citep{ser07}.  The effect of a tangential peculiar velocity has more peculiar signatures. \citet{mo+bi03} noted how the motion of a cluster of galaxies across the line of sight affects observations in weak gravitational lensing regime through distortion of the apparent ellipticity of background galaxies. Even if we have shown that, due to aberration, the distortion matrix can not be written in terms of the second derivatives of a deflection potential, the two approaches still agree on the main point: the effect on the measured background galaxy ellipticities is quadratic in the components of the velocities perpendicular to the line of sight, i.e. proportional to $ v_\perp^2$. Even for a significant peculiar velocity of $\sim 1000 ~\mathrm{km}~\mathrm{s}^{-1}$, the perturbation due to aberration would be $\sim 10^{5}$ times smaller than the main signal. Since the weak lensing signal to noise ratio for detection of a massive cluster with a dispersion velocity of $\sim 1200~\mathrm{km}~\mathrm{s}^{-1} $, obtained through deep observations reaching a background galaxy density of $\sim 30$~gal per arcminute$^2$, is S/N$\sim 15-20$ \citep{sch06}, the measurement of the perturbation is then very challenging. The distortion in the image shape due to aberration should be however taken into account in analyses of flexion, which describes the lowest-order deviation of the lens mapping from its linear expansion and deforms round images into arclets resembling the shape of a banana \citep{go+ba05}.

\section{Final remarks}
\label{sec:conc}

It has been clear for several years that the bending angle by a moving lens carries a factor $\sim (1+z_\mathrm{d})$ with respect to the static case. An insertion of such an angle in an otherwise `static' lens equation would bring to the wrong conclusion that the scaling factor for the Chwolson radius is $\sim (1+z_\mathrm{d}/2)$, in clear disagreement with simple arguments based on aberration of light. In this paper, a lens equation has been derived which can account for a relative motion without biases. The equation has been based on both coordinate deflection in the geodesic motion of light-rays in the rest frame of the lens and aberration in the moving observer's frame. Within this well defined frame-work, the correct pre-factor $\sim (1+z_\mathrm{d})$ for the Einstein ring is restored.

The bending angle by a deflector in motion is usually obtained by applying a Lorentz coordinate transformation to the space-time of the same deflector at rest \citep{fri03a,wu+sp04}. The approach taken in the present paper has been opposite, since I considered the geodesic motion in the Schwarzschild spacetime describing the lens at rest and then interpreted the measurements in the moving frame of the observer, boosted by a Lorentz transformation with respect to the static observer. This made possible to end up with a very simple lens equation written in terms of observable quantities rather than coordinates. As far as weak gravitational fields are concerned, gravity can be put in a
Lorentz-invariant linear form, so that the spacetime of a lens moving with constant speed in the frame of a static observer is the same as the Lorentz-transformed static lens. It is also to be stressed that both methods are based on Lorentz transformations and then assume that accelerations are negligible. In fact, the two methods pick up two different times. The first one consider the velocity of the lens at the time the photon passes by the lens  \citep{fri+al02,fri03a,ser05}, the second one refers to the velocity of the observer at the photon reception. Obviously, when the acceleration is null, such velocities are coincident. Therefore, there is no difference between the observations made by a moving observer on a static lens spacetime or a static observer in a moving lens spacetime. Furthermore, under an astrophysical point of view, the scenario of a static lens and a moving observer can often provide a more immediate representation of the real lensing system, as for the case of lensing by the supermassive black hole in the Galactic center.

The present study has considered aberration effects in the first order of deflection on light rays propagating far from the black hole. However, once we have tested that the standard aberration formula can be applied to the Einstein ring as well as to the bending angle, such formula should likely apply also to relativistic images formed in the strong deflection limit. Then, the relativistic Einstein rings which form near the photon sphere due to light rays winding several times around the black hole suffer the same scaling correction factor of $(1+z_\mathrm{d})$.

A relative motion of the lens affects time delay and redshift of the images too. Even if the lens is in motion, at linear order in velocity the bending angle and the gravitational time delay can still be related by a gradient and it turns out that the time delay by a deflector in motion carries the same pre-factor of $\sim (1+z_\mathrm{d})$ with respect to the standard Shapiro time delay \citep{fri+al02,fri03a,fri03b}. 

Lens motion also affects the observed redhifts of the images \citep{py+bi93,ko+sc99,fri03a,wu+sp04}. The tangential speed of the deflector across the line of sight causes a change in the momentum of the photons so that each image undergoes an additional redshift of $\Delta z = \alpha v_\perp /(1+ v^{(r)})$  \citep{wu+sp04}.  In principle, redshift observations could then be used to determine tangential peculiar velocities by measuring the relative frequency shifts between multiple images of a single strongly lensed background galaxy behind a large cluster \citep{mo+bi03}.

\begin{acknowledgments}
M.S. is supported by the Swiss National Science Foundation and by the Tomalla Foundation.  
\end{acknowledgments}


\end{document}